\documentclass[prb,aps,twocolumn,floatfix,showpacs]{revtex4}
\usepackage{graphicx}

\begin{document}

\title{Statistical significance of fine structure in the frequency spectrum of Aharonov-Bohm conductance oscillations.}

\author{F.E. Meijer$^{1,2}$} \email{f.e.meijer@tnw.tudelft.nl}
\author{A.F. Morpurgo$^1$}
\author{T.M. Klapwijk$^1$}
\affiliation{$^1$Delft University of Technology, Department of
NanoScience, Lorentzweg 1, 2628 CJ Delft, The Netherlands
\\$^2$NTT Basic Research Laboratories, NTT Corporation,
Atsugi-shi, Kanagawa 243-0198, Japan}
\author{J. Nitta$^{1,2}$}
\author{T. Koga$^{1,3}$}
\affiliation{$^1$NTT Basic Research Laboratories, NTT Corporation,
Atsugi-shi, Kanagawa 243-0198, Japan} \affiliation{$^2$CREST-Japan Science and Technology \\
$^3$PRESTO-Japan Science and Technology}

\date{\today}

\begin{abstract}
We discuss a statistical analysis of Aharonov-Bohm conductance
oscillations measured in a two-dimensional ring, in the presence
of Rashba spin-orbit interaction. Measurements performed at
different values of gate voltage are used to calculate the
ensemble-averaged modulus of the Fourier spectrum and, at each
frequency, the standard deviation associated to the average. This
allows us to prove the statistical significance of a splitting
that we observe in the $h/e$ peak of the averaged spectrum. Our
work illustrates in detail the role of sample specific effects on
the frequency spectrum of Aharonov-Bohm conductance oscillations
and it demonstrates how fine structures of a different physical
origin can be discriminated from sample specific features.
\end{abstract}


\maketitle

The investigation of Aharonov-Bohm (AB) conductance oscillations
in mesoscopic devices permits to study different aspects of phase
coherent transport of electrons. One of the aspects that has
recently attracted considerable attention is the effect of the
electron spin \cite{Loss-Stern}. It has been theoretically
predicted that in the presence of spin-orbit interaction (SOI),
the electron spin modifies the properties of AB conductance
oscillations in an observable way \cite{A-LG,Engel-Loss}.

Experimental attempts have been reported in which features
observed in either the envelope function of the AB oscillations or
their Fourier spectrum were attributed to the presence of Rashba
SOI \cite{Morpurgo,Shayegan}. In a few cases \cite{Shayegan} these
claims were based on the interpretation of {\it single}
magneto-conductance measurements. The interpretation of such
experiments is difficult, however, due to the sample specific
nature of the $h/e$ oscillations. In particular, a certain scatter
configuration in the ring might cause features that are similar to
those due to SOI. In the analysis of past experiments this
possibility has not been considered thoroughly.

\begin{figure*}[t]
\begin{center}\leavevmode
\includegraphics[width=0.5\linewidth]{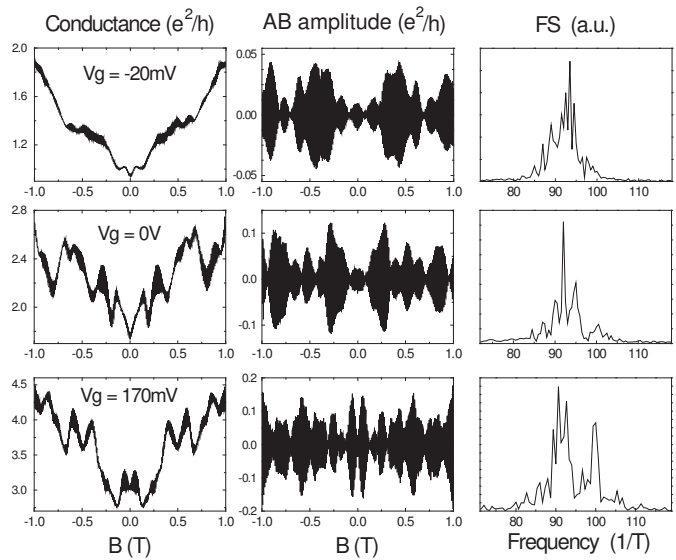}
\caption{The magneto-conductance of a ring measured at different
gate voltage ($V_g$) at 300mK is shown in the left column. The
envelope function of the AB conductance oscillations, obtained by
removing the positive magneto-conductance background, and the
Fourier spectrum are shown in the middle and right column,
respectively. It is apparent that both these quantities depend
strongly on $V_g$, as it is typical for sample specific effects.}
\label{1}
\end{center}
\end{figure*}

In this paper we show experimentally how sample specific effects
in the Fourier spectrum (FS) of the AB oscillations can be {\it
quantifiably} suppressed in a controlled way. In particular, we
perform a statistical analysis of the ensemble averaged FS. At
each frequency, the mean Fourier amplitude and  standard deviation
are calculated. We find features in the averaged FS that are
significantly larger than the standard deviation. These features
can therefore be discriminated from remnant sample specific
effects and their origin attributed to a different physical
phenomenon.

The AB oscillations used in our analysis have been measured in a
two dimensional ring fabricated using an InGaAs-based
heterostructure (Fig. 2, bottom inset), in which Rashba SOI is
particularly strong \cite{Nitta}(Fig. 2, top inset); $\alpha
\approx 0.8 \cdot 10^{-11} $ eVm. The mean radius and width,
respectively 350nm and 180nm, are smaller than the mean free path
($\simeq 1 \mu m$) and transport is quasi-ballistic
\cite{fabrication}. A gate electrode covering the ring permits to
change the Fermi energy as well as the strength of the SOI (the
maximum expected gate induced change is 20\%-30\%) \cite{Nitta}.

The magneto-conductance of the ring ($G(B)$) was measured at
different values of the gate voltage ($V_g$) ranging from -55mV to
195mV at a temperature of 300mK. Three $G(B)$ curves, measured at
different $V_g$, are shown in the left column of Fig.1. Clearly
visible in each $G(B)$ curve are a background increasing with
magnetic field (due to the classical dynamics of the electron in a
laterally confined geometry\cite{Beenakker-VanHouten}), aperiodic
conductance fluctuations, and periodic AB oscillations \cite{WAL}.
The middle column shows the AB oscillations obtained by
subtracting the background from the $G(B)$ curve. Due to the small
period of the oscillations, only the envelope function is visible.
The right column shows the $h/e$ peak in the FS.

Both the envelope function of the AB oscillations and its FS
depend strongly on $V_g$. This strong $V_g$ dependence is expected
and well-known for random interference of electronic waves and
shows that in a single $G(B)$ trace sample specific effects
dominate the behavior of the measured AB oscillations. This
precludes the attribution of a special meaning to any feature
observed in a single measurement, since features appear and
disappear randomly with $V_g$. Therefore, in practice, it is not
possible to draw firm conclusions about the effect of SOI on the
AB oscillations from a single $G(B)$ measurement\cite{Shayegan}.

In order to put in evidence subtle effects possibly present in the
AB oscillations, it is necessary to suppress sample specific
features in a controlled way. This can be achieved by studying an
\textit{ensemble of measurements}\cite{Morpurgo}, i.e. by
averaging the modulus of the Fourier spectrum, $|G(\nu)|$, over
different scatterer configuration. The ensemble averaged FS,
$\langle |G(\nu)| \rangle$, is expected to be a smooth function,
peaked at the frequency that corresponds to the mean radius of the
ring. Superimposed on top of this smooth function, spin (or other)
effects may show up as a splitting or another well-defined
structure.

Theoretically it has been proven that a (sufficiently large)
change in Fermi energy is equivalent to a complete change in
impurity configuration, insofar as the conductance oscillations
are concerned \cite{Altshuler}. For this reason we study the
statistical properties of quantities averaged over an ensemble
that consists of $G(B)$ curves measured at different gate voltage.

To show experimentally the statistical independence of $G(B)$
curves measured at different values of $V_g$ we calculate the mean
amplitude of the $h/e$ oscillations upon averaging an increasing
number $N$ of $G(B)$ traces. The $h/e$ oscillation amplitude in
the averaged $G(B)$ is expected to be suppressed as $1/\sqrt{N}$,
provided that the curves are statistically
independent\cite{Washburn-Webb}. Fig. 2 shows that this is indeed
the behavior observed experimentally \cite{normalization}. We have
checked that also the aperiodic conductance fluctuations exhibit
the same behavior, i.e. their amplitude decreases as $1/\sqrt{N}$
as well.

We note that in order to acquire the largest possible number of
independent $G(B)$ curves, and therefore to obtain the largest
suppression of sample specific effects, the entire range of $V_g$
studied is rather large. This results in a sizeable (up to roughly
a factor of 2) change of electron density and possibly also in a
change of SOI strength, which may affect the shape of the averaged
modulus of the FS, $\langle |G(\nu)| \rangle$. For this reason, we
will first discuss the average over the whole $V_g$ range, and
then compare it to the average over a smaller $V_g$ range. As we
will show, the final results are similar for the two different
$V_g$ ranges, which suggests that the precise extension of the
$V_g$ range used in the experiments is not very critical.

Figure 3 shows a typical FS of a single $G(B)$ trace (upper
graph), and the result of two different kinds of ensemble
averaging, using the total set of 49 $G(B)$ curves measured from
$V_g$= -55mV to 195mV. Specifically, the middle graph is a plot of
the modulus of the FS of the average magneto-conductance, obtained
by first averaging the magneto-conductance curves and then
calculating the modulus of the FS ($\equiv$ $|\langle G(\nu)
\rangle|$). This quantity has been studied extensively in the
past, and was shown to result in a suppression the $h/e$
oscillation amplitude\cite{Washburn-Webb}. For this reason, the
{\it relative} size of the sample specific structure in the FS
does not decrease upon averaging, as it is apparent from the
comparison of the upper and middle graph in Fig. 3. Therefore,
this way of ensemble averaging does not allow the observation of
subtle features possibly present in the $h/e$ peak in the FS.

\begin{figure}[t]
\begin{center}\leavevmode
\includegraphics[width=0.85\linewidth]{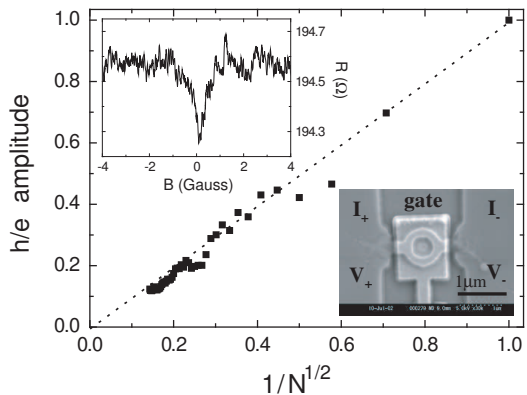}
\caption{Mean amplitude of the $h/e$ oscillations in the averaged
magneto-conductance, as a function of the square root of the
number $N$ of curves used in calculating the average. The mean
amplitude of the $h/e$ oscillations is calculated by integrating
the modulus of the FS over the width of the $h/e$ peak. The insets
show the sample used in our investigations (bottom right) and the
low-field magnetoresistance measured in a Hall bar made out of the
same heterostructure used for our ring (top left). The resistance
dip around zero field is due to weak anti-localization, which
indicates the presence of a strong Rashba spin-orbit
interaction.}\label{2}
\end{center}

\end{figure}
\begin{figure}[t]
\begin{center}\leavevmode
\includegraphics[width=0.8\linewidth]{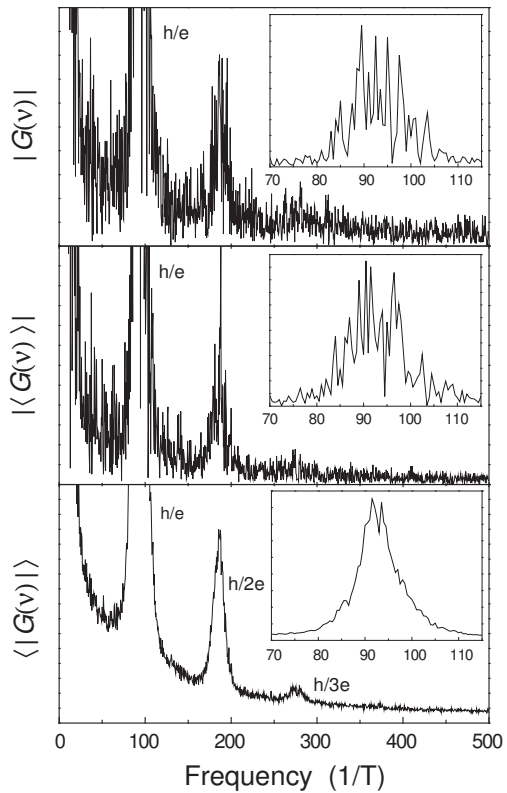}
\caption{The random features present in the modulus of the FS of a
single mangeto-resistance curve (upper graph) are {\it not}
suppressed in the modulus of the FS of the averaged $G(B)$,
$|\langle G(\nu) \rangle |$ (middle graph). However, random
features are clearly suppressed in the averaged modulus of the FS,
$\langle | G(\nu) | \rangle$ shown in the bottom graph (in both
the middle and in the bottom graph, the average is taken over
$N=49$ curves measured at different gate voltages). Note the much
enhanced visibility of the $h/3e$ peak in $\langle | G(\nu) |
\rangle$. The insets show the effect of samples specific features
in the frequency range corresponding to the $h/e$ peak. Also here
the difference between the middle and bottom graph is obvious and
a splitting is clearly visible in the $h/e$ peak of in $\langle |
G(\nu) | \rangle$ (note, in the main panel, that a small splitting
is also present on top of the $h/3e$ peak)} \label{3}
\end{center}
\end{figure}

The bottom graph in Fig. 3 shows the ensemble average of the {\it
absolute} value of the FS, $\langle |G(\nu)| \rangle$. This way of
averaging does not suppress the $h/e$ oscillations, since the
phase information is discarded by taking the modulus of the FS of
individual $G(B)$ traces before performing the average. This
procedure results in a suppression of random features in the
averaged FS, over the whole frequency range, as it is obvious from
Fig. 3.

It was argued in ref [4] that this way of ensemble averaging
provides information that is not easily accessible otherwise. The
enhanced visibility of the third harmonic of the AB oscillations
gives a direct experimental demonstration of this statement. This
$h/3e$ peak is not observable in the FS of any single $G(B)$ curve
nor in the FS of the averaged magneto-conductance, but it is
clearly visible in the averaged modulus of the FS, $\langle
|G(\nu)| \rangle$.

The insets in Fig.3 zoom in on the $h/e$ peak. It is apparent
again that, in comparison to the upper two graphs, the sample
specific "noise" is largely suppressed in the averaged modulus of
the FS (bottom graph). In this quantity, a small splitting remains
visible in the averaged $h/e$ peak. Note that also the $h/3e$ peak
in the bottom main figure shows a similar structure.

Without any further analysis, it is difficult to conclude what the
origin of these features is. Specifically, this is because
experimentally we only average over a finite number of scatter
configurations so that the splitting may simply be some remnant
sample specific structure. Only by quantifying the magnitude of
these remnant features we can conclude if the splitting has
physical significance.

We quantify the size of the remnant sample specific structure in
terms of a frequency dependent standard deviation. This is
obtained from the same set of $N$ curves that we use to calculate
the average modulus of the FS, $\langle |G(\nu)| \rangle$. At each
fixed frequency $\nu$, this set of curves corresponds to a set of
$N$ values of which we calculate the standard deviation
$\sigma_{dis}(\nu)$. The statistical error associated with the
average modulus of the FS at frequency $\nu$ is then
$\sigma_{mean}(\nu) = \sigma_{dis}(\nu)/\sqrt{N}$ (central limit
theorem). For an ideal ensemble average $N = \infty$ and
$\sigma_{mean}(\nu) = 0$, i.e. sample specific effects are
completely suppressed. However, if $N$ is finite,
$\sigma_{mean}(\nu)$ is also finite.

The upper panel of Fig. 4 shows the $h/e$ peak of the averaged
modulus of the FS and, for each frequency, $\sigma_{mean}(\nu)$,
plotted as an error bar. In the main panel the average has been
performed on $N=16$ curves (with $V_g$ ranging from -55mV to 95mV)
and in the inset on $N=49$ ($V_g$ ranging from -55mV to 195mV). In
both cases the size of the splitting is 3 to 4 times larger than
$\sigma_{mean}$. The splitting is therefore statistically
significant, and it is {\it not} due to remnant sample specific
structure\cite{analysis}. The size of any other structure observed
in these averaged quantities is too small compared to
$\sigma_{mean}(\nu)$ to exclude remnant sample specific effects as
their origin. This is also true for the splitting in the $h/3e$
peak visible in the bottom panel Fig. 3.

\begin{figure}[t]
\begin{center}\leavevmode
\includegraphics[width=0.9\linewidth]{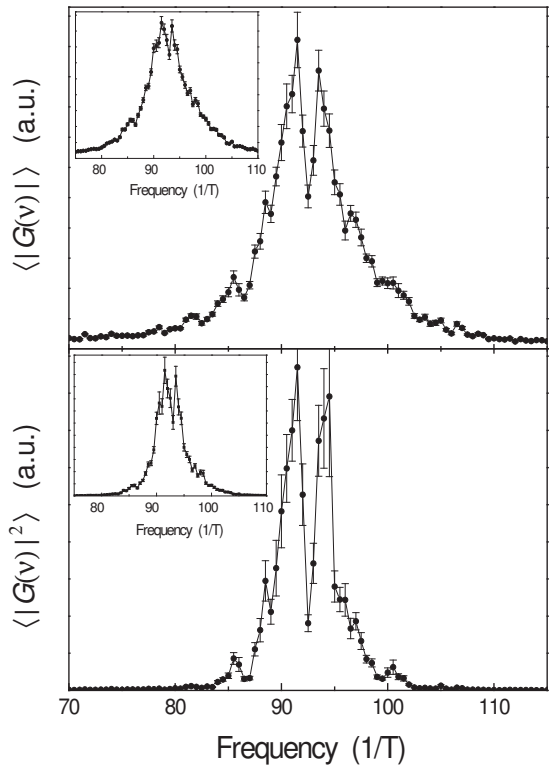}
\caption{The upper graph shows the $h/e$ peak in the averaged
modulus of the Fourier spectrum and the associated statistical
error $\sigma_{mean}$. The average is performed over $N=16$ (main
figure) and $N=49$ (inset) curves, measured at different $V_g$.
The bottom graph shows the same statistical analysis for the power
spectrum. The observed splittings are significantly larger than
the corresponding $\sigma$ and can therefore be discriminated from
remnant sample specific features and attributed to a different
physical phenomenon.} \label{4}
\end{center}
\end{figure}

We have also performed the same statistical analysis for the
Fourier {\it power} spectrum\cite{powerspectrum}. The ensemble
averaged Fourier power, $\langle |G(\nu)|^2 \rangle$, and the
associated statistical error, $\sigma_{power} (\nu)$, are shown in
the bottom panel of Fig. 4 for $N=16$ (main figure) and $N=49$
(inset). Also here a statistically significant splitting is
present in both cases.

The results shown in Fig.4 indicate that the presence of a
statistically significant splitting in the $h/e$ peak is a robust
feature. It does not depend on the precise gate voltage range used
in our analysis, nor on the specific quantity analyzed, i.e. the
modulus of the FS or the power spectrum. The result is also robust
against different procedures used to calculate the Fourier spectra
from the experimental data. We have analyzed our data with and
without removing the background in the magneto-conductance in
different ways and using different kinds of windowing procedures.
In all cases the final results shows a similar, statistically
significant splitting\cite{checks}.

Our results confirm the conclusion of Ref. [4], namely that the
$h/e$ peak in the averaged FS is split in the presence of SOI, and
put it on firmer grounds for two reasons. First, because the
statistical analysis of the significance of the splitting had not
been previously performed. Second, because the same effect is
observed in a different material system.

The size of the splitting measured here is roughly 3/T, smaller
than the one found in Ref. [4] (approximately 12/T). A direct
comparison of the two results is however difficult, because the
heterostructures used in the two experiments are different, as
well as the radius of the rings (0.35 $\mu m$ versus 1.05 $\mu
m$). It was argued in Ref [4] that the splitting may be a
manifestation of the geometrical Berry phase. Subsequent
theoretical works seem to consistently conclude that the effect of
Berry phase alone is too small to account for the magnitude of the
previously observed splitting\cite{averaged FS}. This is also true
for the magnitude of the splitting observed here, whose precise
origin remains to be identified. More experimental work is needed
to discriminate between different possible mechanisms capable of
accounting for the experimental results (such as SOI\cite{averaged
FS,Engel-Loss} or Zeeman\cite{Engel-Loss}). The work presented in
this paper demonstrates a well-defined experimental procedure on
which future experiments can be based.

In conclusion we have discussed in detail the statistical
properties of experimentally measured Aharonov-Bohm conductance
oscillations and shown how random sample specific effects can be
suppressed in a {\it quantifiable} way, up to a level that permits
to reveal features of a different physical origin.

We would like to acknowledge Dr. Y. Nazarov and Dr. Y. Blanter for
discussions and Dr. H. Takayanagi and the Stichting FOM for
support. FEM is grateful to NTT for hospitality and financial
support, which made this work possible. The work of AFM is part of
the NWO Vernieuwingsimpuls 2000 program.


\begin{thebibliography}{9}
\bibitem{WAL} Signs of weak anti-localization due to SOI are not visible in these $G(B)$ curves,
since the associated magnetic field scale is well within one AB
oscillation (see inset of Fig.2).
\bibitem{normalization} The mean $h/e$ oscillation amplitude was normalized to unity for each $G(B)$
curve in order to ensure that each curve contributes equally to
the average. We have checked that this normalization procedure
does not affect our final results, presented in Fig.4: The same
features are obtained using non-normalized $G(B)$ curves.
\bibitem{analysis} Note that although more traces are used in the average over the larger $V_g$ range
the size of the splitting, measured in terms of $\sigma$, does not
increase. This is possibly due to some averaging-out of SOI
effects with large $V_g$ ranges.
\bibitem{powerspectrum} The (ensemble averaged) Fourier power
spectrum is equal to the Fourier transform of the (ensemble
averaged) magnetic correlation function $\langle C(\Delta B)
\rangle$, where $C(\Delta B)=\int G(B) G(B+\Delta B)~dB$.
\bibitem{checks} The classical background and "edge effects" were found
to result predominantly in low frequency components in the FS, far
below the $h/e$ peak. As a consequence, also the shape of the
background (roughly above 25/T) is not/hardly affected.
\end{thebibliography}
\end{document}